\documentclass[twocolumn,prl,aps,psfig,showpacs,preprintnumbersm,superscriptaddress]{revtex4}
\usepackage{graphicx}
\usepackage{epstopdf}
\usepackage{float}
\usepackage{color}
\usepackage{amsmath}

\begin{document}
\title{
Dominant in-plane symmetric elastoresistance in CsFe$_2$As$_2$
}

\author{P.~Wiecki}
\affiliation{Karlsruhe Institute of Technology, Institute for Quantum Materials and Technologies, 76021 Karlsruhe, Germany}
\author{A.-A.~Haghighirad}
\affiliation{Karlsruhe Institute of Technology, Institute for Quantum Materials and Technologies, 76021 Karlsruhe, Germany}
\author{F.~Weber}
\affiliation{Karlsruhe Institute of Technology, Institute for Quantum Materials and Technologies, 76021 Karlsruhe, Germany}
\author{M.~Merz}
\affiliation{Karlsruhe Institute of Technology, Institute for Quantum Materials and Technologies, 76021 Karlsruhe, Germany}
\author{R.~Heid}
\affiliation{Karlsruhe Institute of Technology, Institute for Quantum Materials and Technologies, 76021 Karlsruhe, Germany}
\author{A.~E.~B\"{o}hmer}
\affiliation{Karlsruhe Institute of Technology, Institute for Quantum Materials and Technologies, 76021 Karlsruhe, Germany}
\date{\today}

\begin{abstract}
We study the elastoresistance of the highly correlated material CsFe$_2$As$_2$ in all symmetry channels. Neutralizing its thermal expansion by means of a piezoelectric-based strain cell is demonstrated to be essential. The elastoresistance response in the in-plane symmetric channel is found to be large, while the response in the symmetry-breaking channels is weaker and provides no evidence for a divergent nematic susceptibility.
 Rather, our results can be interpreted naturally within the framework of a coherence-incoherence crossover, where the low-temperature coherent state is sensitively tuned by the in-plane atomic distances.
\end{abstract}

\maketitle

In recent years, strain has been increasingly recognized as a valuable parameter for tuning the electronic properties of strongly-correlated quantum materials \cite{Hicks2014a,Steppke2017,Kim2018}. In contrast to hydrostatic pressure, strain application offers the ability to probe distortions of varying symmetries and to expand lattice parameters as well as compress them.
The strain dependence of the electrical resistivity, known as elastoresistance, has emerged as a simple and powerful tool to study a wide class of materials \cite{Riggs2015,Ishida2019b,Rosenberg2019,Jo2019}. 
For example, in the iron-based superconductors, elastoresistance measurements
have been crucial to the understanding of the electronically-driven nematic state and its relation to superconductivity \cite{Chu2012,Watson2015,Kuo2016}. 
Due to the intrinsic resistivity anisotropy of the nematic state, the resistance changes induced by anisotropic strains can be used as a measure of the nematic susceptibility \cite{Chu2012}. Nematicity, the breaking of rotational symmetry via electronic interactions, has indeed emerged as a key research area in correlated electron systems \cite{Borzi2007,Hinkov2008,Fernandes2014,Wu2017,Ronning2017,Yonezawa2017}.

The iron-based superconductor CsFe$_2$As$_2$ hosts 
a unique strongly-correlated electronic state \cite{Hardy2013,Eilers2016,Hardy2016}. These correlations have been shown to grow with hole-doping in the isostructural series Ba$_{1-x}$K$_x$Fe$_2$As$_2$ up to its endmember KFe$_2$As$_2$ and then grow further upon isoelectronic substitution of K by the larger Rb and Cs ions, where the highest values are reached\cite{Hardy2016}. The Sommerfeld coefficient, $\gamma$, which characterizes such correlations, 
amounts to 
180 mJ/mol/K$^2$ in CsFe$_2$As$_2$, a value comparable to some heavy-fermion compounds. Furthermore, resistivity, magnetization, NMR and thermal expansivity measurements all found evidence for a crossover between a low-temperature heavy Fermi-liquid regime and a high
temperature regime of reduced quasiparticle coherence \cite{Hardy2013,Hardy2016,Haule2009,Wu2016,Zhang2018,Zhao2018,Wiecki2018}. These properties are signatures of heavy-Fermion behavior, which, uncommonly, emerges in these $d$-electron systems. It is believed to arise from proximity to an orbital-selective Mott phase, 
resulting from
 large Hund's coupling \cite{Medici2014}. Similar to the chemical pressure from substituting larger ions, physical in-plane-symmetric tensile strain may further amplify this behavior, as can be inferred from the thermal expansivity data\cite{Eilers2016,Hardy2016}, as well as from LDA+DMFT calculations in a fictitious ``stretched" CsFe$_2$As$_2$ system\cite{Backes2015}. 

Recently, a novel electronic nematicity has been proposed in CsFe$_2$As$_2$ and RbFe$_2$As$_2$ \cite{Li2016,Ishida2019,Moroni2019}. 
Elastoresistance measurements reported that in these materials, the nematic direction 
is along [100] (in tetragonal notation), in contrast to the [110] direction typical for undoped iron-based superconductors such as BaFe$_2$As$_2$.
A unique electronic $XY$-nematic state was proposed in the crossover region in the Ba$_{1-x}$Rb$_x$Fe$_2$As$_2$ series  \cite{Ishida2019}. Furthermore, a temperature-driven transition to an ordered nematic state in RbFe$_2$As$_2$ was proposed on the basis of elastoresistance. STM measurements on RbFe$_2$As$_2$ have found a symmetry breaking in the electronic structure and magnetic vortex shapes consistent with such an ordered nematic state \cite{Liu2019}. Nevertheless, such a phase transition has not yet been confirmed by thermodynamic probes.
Several theoretical scenarios for the novel nematicity in these strongly-correlated iron-based systems have been proposed \cite{Onari2019,Wang2019,Borisov2019}.

These recent results ask for a close examination of the relative effect of strains in all different symmetry channels on equal footing. 
Furthermore, as a direct consequence of its strain-dependent correlations, CsFe$_2$As$_2$ has an unusually large thermal expansivity\cite{Hardy2016}, requiring 
an experimental environment where the sample strain can be controlled over a large range.
Here, we present differential elastoresistance measurements of CsFe$_2$As$_2$ obtained in a piezo-based strain cell, which allows for full control of the strain state of the sample. 
We find that this material is primarily sensitive to in-plane symmetric strains 
and does not exhibit a diverging nematic susceptibility. 
These results show that the orbital-selective Mott physics which dominate in this system are sensitively tuned by the average in-plane interatomic distances but do not entail nematicity. 

\begin{figure*}[t]
\centering
\includegraphics[width=\textwidth]{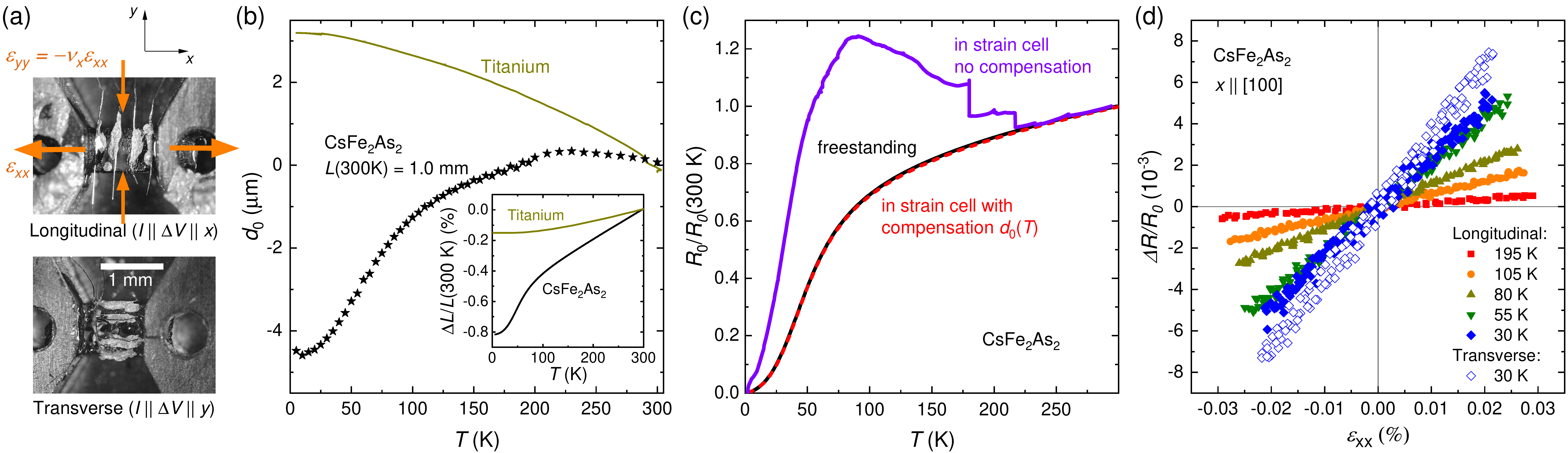}
\caption{(a) Photograph of the experimental setup for transverse and longitudinal elastoresistance measurements with the direction of the applied strains $\varepsilon_{xx}$ and $\varepsilon_{yy}$ indicated.
(b) Black stars: 
Example of the displacement $d_0(T)$ applied to keep a 1.0 mm-long crystal in an unstrained state.
Dark yellow line: Zero-strain displacement for a Titanium reference sample. 
Inset: Change in length as a function of temperature for Titanium and CsFe$_2$As$_2$ (both freestanding).
(c) Unstrained resistance $R_0$ of a freestanding crystal (black), and mounted in the strain cell
with (red) and without (purple) active compensation for the thermal expansion mismatch. 
(d) Relative change in resistance $\Delta R/R_0$ as a function of applied strain $\varepsilon_{xx}||[100]$ in longitudinal (full symbols) and transverse (empty symbols) contact configurations. 
}
\label{fig:setup}
\end{figure*}

Single crystals of CsFe$_2$As$_2$ were grown from solution with a Cs-rich self-flux in an alumina crucible sealed in a stainless steel container. Samples were carefully characterized by single-crystal x-ray refinement \cite{supp}.
Elastoresistance measurements were performed in a commercial strain cell (Razorbill Instruments CS-100). In such a cell, the sample is fixed and suspended between two movable sample plates, the distance between which is controlled by  piezoelectric stacks \cite{Hicks2014}.
The samples were cut with edges along the tetragonal in-plane directions [100] or [110], with typical dimensions of
2.0$\times$0.8$\times$0.07 mm. The samples were fixed in the strain cell using 5-minute epoxy (DevCon) with the effective strained sample length, $L(300K)$, typically 1.1 mm (Fig. \ref{fig:setup}(a)). 
The epoxy thickness was controlled to be 50 $\mu$m below the sample and $\sim$30 $\mu$m above the sample.
The sample strain was measured via a built-in capacitive displacement sensor.  
Resistance was measured using a four-point method on a Lake Shore 372 resistance bridge. Two different configurations of four contacts were used so current could be applied either longitudinal or transverse to the stress axis, defined as the $x$ axis (Fig. \ref{fig:setup}(a)). Longitudinal and transverse elastoresistance measurements were performed on separate crystals from same batch. 
We also conducted complementary $x$-ray diffraction (XRD) measurements on 20-50 $\mu$m thick samples glued to glass substrates
with the same 5-minute epoxy in a Huber 4-circle diffractometer using Mo {\it K}$_\alpha$ radiation. 
Scattered x-rays were detected by a VORTEX silicon-drift point detector. 
The $150\mu$m thickness of the glass allowed for XRD in transmission geometry. 

The large thermal expansion mismatch between CsFe$_2$As$_2$ and the Titanium body of the strain cell (see inset of Fig. \ref{fig:setup}(b)) could apply a large tensile strain of up to $\sim$0.7$\%$ to the sample. Indeed, if this extreme tension is not compensated, the samples mounted in the strain cell were found to crack on cooling, resulting in dramatic changes in the resistance (Fig. \ref{fig:setup}(c)). To keep the sample in an unstrained state, we progressively reduce the distance between the sample mounting plates by an amount $d_0(T)$ as the sample is cooled. 
In order to determine this zero-strain displacement $d_0(T)$, we first measure the zero strain displacement for a Titanium sample using a stiff calibration piece\cite{Ikeda2018,Razorbill}, see Fig. \ref{fig:setup}(b). The zero-strain displacement for the sample of interest is then calculated based on its known thermal expansion difference with Titanium 
(see \cite{supp} for details). Figure \ref{fig:setup}(b) shows the applied $d_0(T)$ for one particular measurement of a CsFe$_2$As$_2$ sample with $L(300K)=1$ mm as an example. This compensation procedure was successful in preventing sample cracking and maintaining it near its freestanding resistance (Fig. \ref{fig:setup}(c)). 
Elastoresistance is finally measured by performing a small displacement sweep about $d_0(T)$, with the strain defined as $\varepsilon_{xx}=(d-d_0(T))/L(300K)$, where $d$ is the applied total displacement.

\begin{figure}[t]
\centering
\includegraphics[width=\columnwidth]{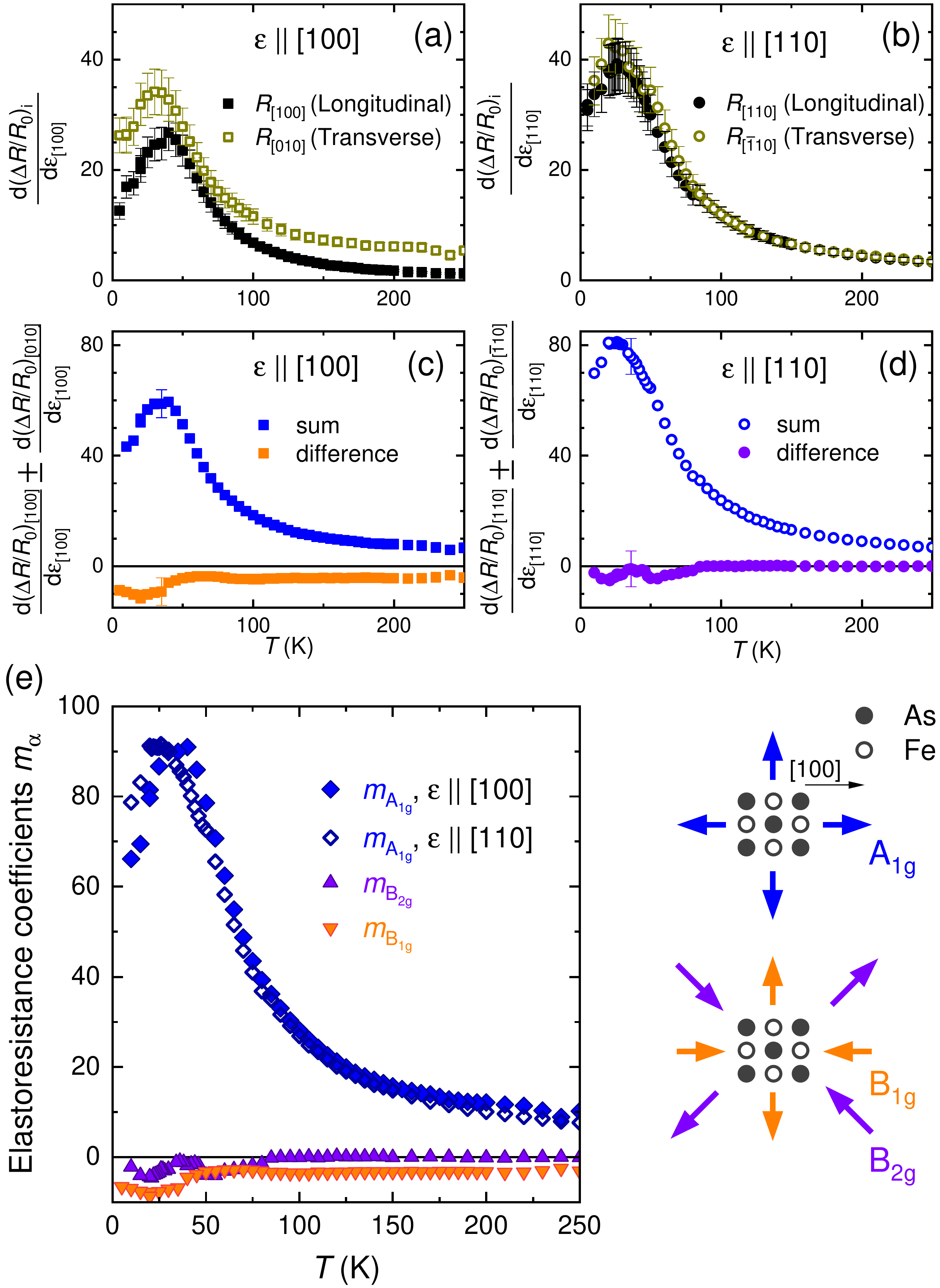}
\caption{Slope $d(R_{ii}/R_0)/d\varepsilon_{xx}$ ($i=x,y$) as a function of temperature for (a) [100] $\| x$  and (b) [110]$ \| x$ (b).
Sum and difference of the longitudinal ($R_{xx}$) and transverse ($R_{yy}$) slopes for (c) [100]$ \| x$  and (d) [110]$ \| x$.
(e) Elastoresistance coefficients for strains belonging to  different irreducible representations, along with their definitions represented by schematic drawings. 
}
\label{fig:finaldata}
\end{figure}

The fractional change of resistance $\Delta R/R_0$, as a function of applied strain is presented in Fig. \ref{fig:setup}(d) for selected temperatures. 
All curves are linear within the strain range investigated ($\pm0.03\%$).
Testing over much larger strain scales of more than 0.2\%, made possible by our use of a strain cell, reveals only a slight non-linearity \cite{supp}.
We conducted transverse and longitudinal measurements for strain applied along both [100] and [110]. For all four configurations, the slope $d(\Delta R/R_0)/d\varepsilon$ is plotted as a function of temperature in Figs. \ref{fig:finaldata}(a),(b). 
Here, we show our estimates of the systematic error, obtained by conducting measurements with the same configuration on multiple crystals. 
We observe similar behavior for all configurations, featuring 
a prominent increase on cooling and a broad peak around 30 K. 
Furthermore, in all configurations, the slope has the same positive sign. This is in clear contrast to materials such as Ba(Fe,Co)$_2$As$_2$, where transverse and longitudinal elastoresistance data have opposite sign due to a dominating nematic response\cite{Kuo2013}.

 We now extract the elastoresistance coefficients $m_\alpha$, which relate the changes in resistance to strains in the corresponding symmetry channels, $(\Delta R/R_0)_\alpha=m_\alpha\varepsilon_\alpha$. Here, $\alpha$ represents the irreducible representation of the tetragonal $D_{4h}$ point group, see the 
schematic illustrations in Fig. \ref{fig:finaldata}(e).
The relevant strain components are given by $\varepsilon_{A_{1g}}=\frac{1}{2}(\varepsilon_{[100]}+\varepsilon_{[010]})=\frac{1}{2}(\varepsilon_{[110]}+\varepsilon_{[\bar{1}10]})$, $\varepsilon_{B_{1g}}=\frac{1}{2}(\varepsilon_{[100]}-\varepsilon_{[010]})$ and $\varepsilon_{B_{2g}}=\frac{1}{2}(\varepsilon_{[110]}-\varepsilon_{[\bar{1}10]})$, where we use
the simplified notation $\varepsilon_x\equiv\varepsilon_{xx}$  for the diagonal strain components \cite{Palmstrom2017,supp}.

When uniaxial strain is applied along $[100]$ (or $[110]$), the sample experiences
$A_{1g}+B_{1g}$ (or $A_{1g}+B_{2g}$) strain. The sum of the longitudinal and transverse resistance change is proportional to $(\Delta R/R_0)_{A_{1g}}$, while the difference is proportional to $(\Delta R/R_0)_{B_{1g}}$ [or $(\Delta R/R_0)_{B_{2g}}$] \cite{Palmstrom2017,supp}, see Fig. \ref{fig:finaldata}(c),(d).
The sample's Poisson ratio $\nu$ relates the strain along different directions, $\nu_{[100]}=-\varepsilon_{[010]}/\varepsilon_{[100]}$, and $\nu_{[110]}=-\varepsilon_{[\bar{1}10]}/\varepsilon_{[110]}$, with $\nu_{[100]}\neq\nu_{[110]}$, in general. Finally, the elastoresistance coefficients are expressed in terms of experimental quantities as,
\begin{subequations}
\begin{align}
m_{A_{1g}}&=\frac{1}{1-\nu_{[100]}}\left[\frac{d(\Delta R/R_0)_{[100]}}{d\varepsilon_{[100]}}+\frac{d(\Delta R/R)_{[010]}}{d\varepsilon_{[100]}}\right]\nonumber\\
&=\frac{1}{1-\nu_{[110]}}\left[\frac{d(\Delta R/R_0)_{[110]}}{d\varepsilon_{[110]}}+\frac{d(\Delta R/R)_{[\bar{1}10]}}{d\varepsilon_{[110]}}\right]\nonumber\\
m_{B_{1g}}&=\frac{1}{1+\nu_{[100]}}\left[\frac{d(\Delta R/R_0)_{[100]}}{d\varepsilon_{[100]}}-\frac{d(\Delta R/R)_{[010]}}{d\varepsilon_{[100]}}\right]\nonumber\\
m_{B_{2g}}&=\frac{1}{1+\nu_{[110]}}\left[\frac{d(\Delta R/R_0)_{[110]}}{d\varepsilon_{[110]}}-\frac{d(\Delta R/R)_{[\bar{1}10]}}{d\varepsilon_{[110]}}\right]\nonumber.
\end{align}
\end{subequations}

The sample Poisson ratios were calculated from the elastic constants of the material. Estimates of the elastic constants were extracted from ab-initio calculations via the method of long waves \cite{Eilers2016,supp}. We find that $\nu_{[100]}=0.346$ and $\nu_{[110]}=0.113$, which we take to be temperature independent.

The extracted $m_{A_{1g}}$, $m_{B_{1g}}$ and $m_{B_{2g}}$ are shown in Fig. \ref{fig:finaldata}(e). The in-plane symmetric $A_{1g}$ response clearly dominates the elastoresistance in CsFe$_2$As$_2$ and has a distinct temperature dependence. Note that the $m_{A_{1g}}$ data agree whether the coefficient is evaluated from the [100] or [110] elastoresistance. 
This agreement between independent measurements 
suggests that our data are robust to systematic error.
The $m_{B_{1g}}$ and $m_{B_{2g}}$ coefficients are much smaller than $m_{A_{1g}}$ and nearly independent of temperature. As $m_{B_{1g}}$ and $m_{B_{2g}}$ are a measure of nematic susceptibility in the respective channels \cite{Kuo2016}, we see no evidence for a nematic instability in CsFe$_2$As$_2$.

\begin{figure}[t]
\centering
\includegraphics[width=\columnwidth]{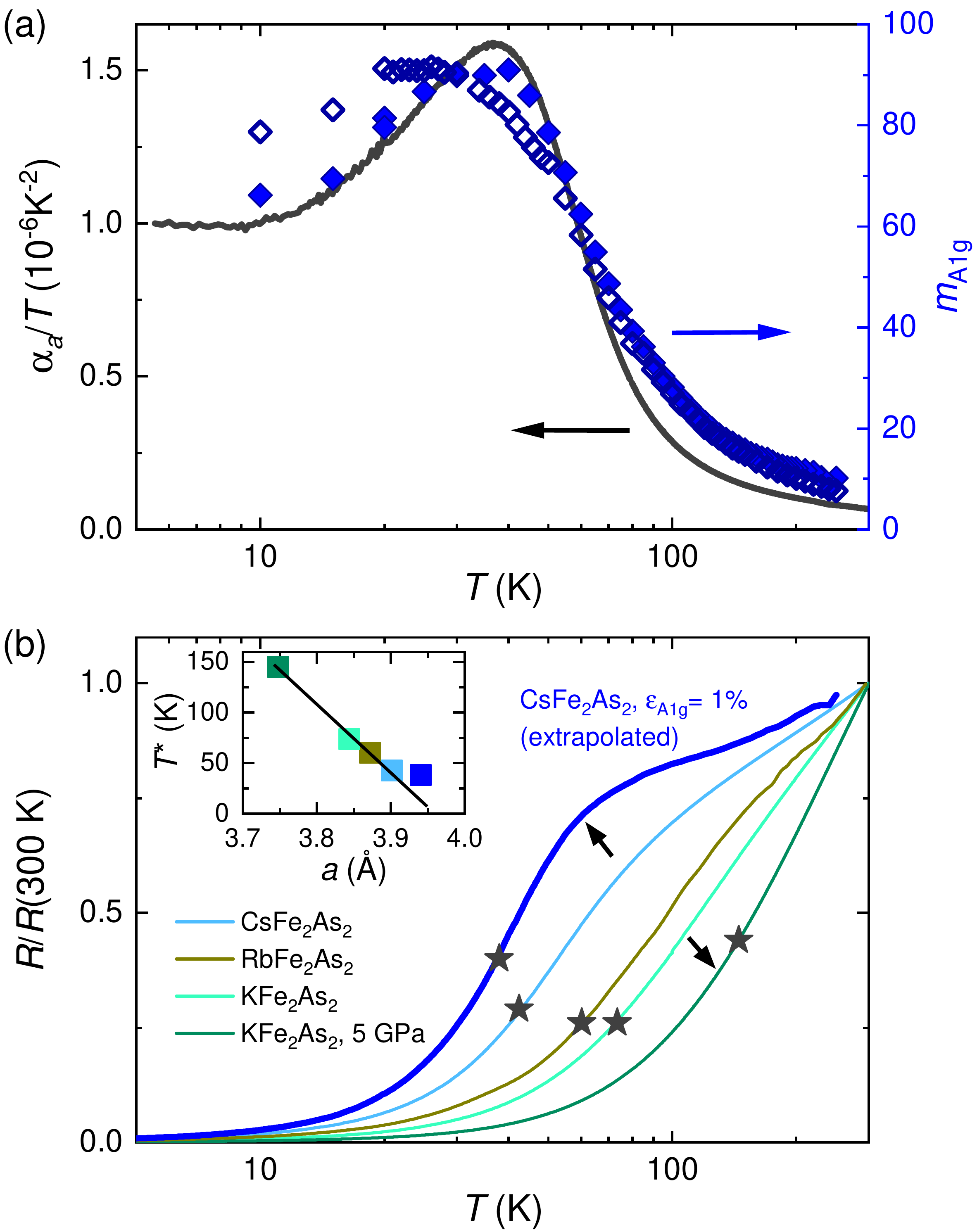}
\caption{
(a) Comparison of the $A_{1g}$ elastoresistance coefficient with the in-plane coefficient of thermal expansion $\alpha_a/T$ on scaled axes. (b) The resistance of CsFe$_2$As$_2$ under $1\%$ biaxial tension, inferred from the linear $A_{1g}$ elastoresistance coefficient.
The resistance of freestanding AFe$_2$As$_2$ (A = K\cite{Taufour2014,Wiecki2018}, Rb\cite{Wu2016}, Cs) crystals, as well as KFe$_2$As$_2$ under 5 GPa hydrostatic pressure \cite{Taufour2014,Wiecki2018}
are also shown for comparison. Stars represent the inflection point of the resistance, which is a
measure of the coherence-incoherence crossover temperature $T^*$.
Inset: $T^*$ as a function of the lattice parameter $a$ (structural data from \cite{Eilers2016,Tafti2014}).
}
\label{fig:discussion}
\end{figure}

The temperature dependence of $m_{A_{1g}}$ compares well with that of the in-plane coefficient of thermal expansion divided by temperature, $\alpha_a/T$, as shown in Fig. \ref{fig:discussion}(a).
 The increase on cooling and broad maximum around 30 K are characteristic of the coherence-incoherence crossover \cite{Hardy2016}.
 Why do $m_{A_{1g}}$ and $\alpha_a/T$ have a similar temperature dependence?
From thermodynamic relations, the linear thermal expansivity is a measure of the (symmetry-conserving) in-plane pressure derivative of the electronic entropy. Since the elastoresistance coefficient $m_{A_{1g}}$ measures the in-plane symmetric strain derivative of the resistance, their proportionality provides evidence that the resistance, being dominated by electronic scattering, is a good gauge for the electronic entropy in this system. 

Indeed, the resistance is known to be sensitive to the coherence-incoherence crossover in the AFe$_2$As$_2$ (A = K, Rb, Cs) compounds \cite{Wu2016,Wiecki2018}.
The residual resistivity reaches very high values of near to 1000 in these compounds, due to the high scattering rate in the high-temperature incoherent state \cite{Hardy2013}.
The inflection point of the resistance provides an estimate of the coherence-incoherence crossover temperature $T^*$. In these samples, lower $T^*$ is associated with increases in both the Sommerfeld coefficient 
$\gamma$ and $\partial\gamma/\partial p$ \cite{Hardy2016},
and increased spin-fluctuations \cite{Wu2016,Zhang2018,Wiecki2018}. As the resistance data in Fig. \ref{fig:discussion}(b) shows, $T^*$ decreases from KFe$_2$As$_2$ to RbFe$_2$As$_2$ to CsFe$_2$As$_2$ as the lattice parameters expand [inset in Fig. \ref{fig:discussion}(b)]. 
Consistently, hydrostatic pressure suppresses the correlations and increases $T^*$, as demonstrated by the example of KFe$_2$As$_2$ under 5 GPa. 
To visualize the effect of biaxial tension, we use our $m_{A_{1g}}$ data and extrapolate the resistance of CsFe$_2$As$_2$ linearly to 1\% biaxial tension. Clearly, this results in the most correlated behavior with the lowest $T^*$. 

\begin{figure}[tb]
\centering
\includegraphics[width=\columnwidth]{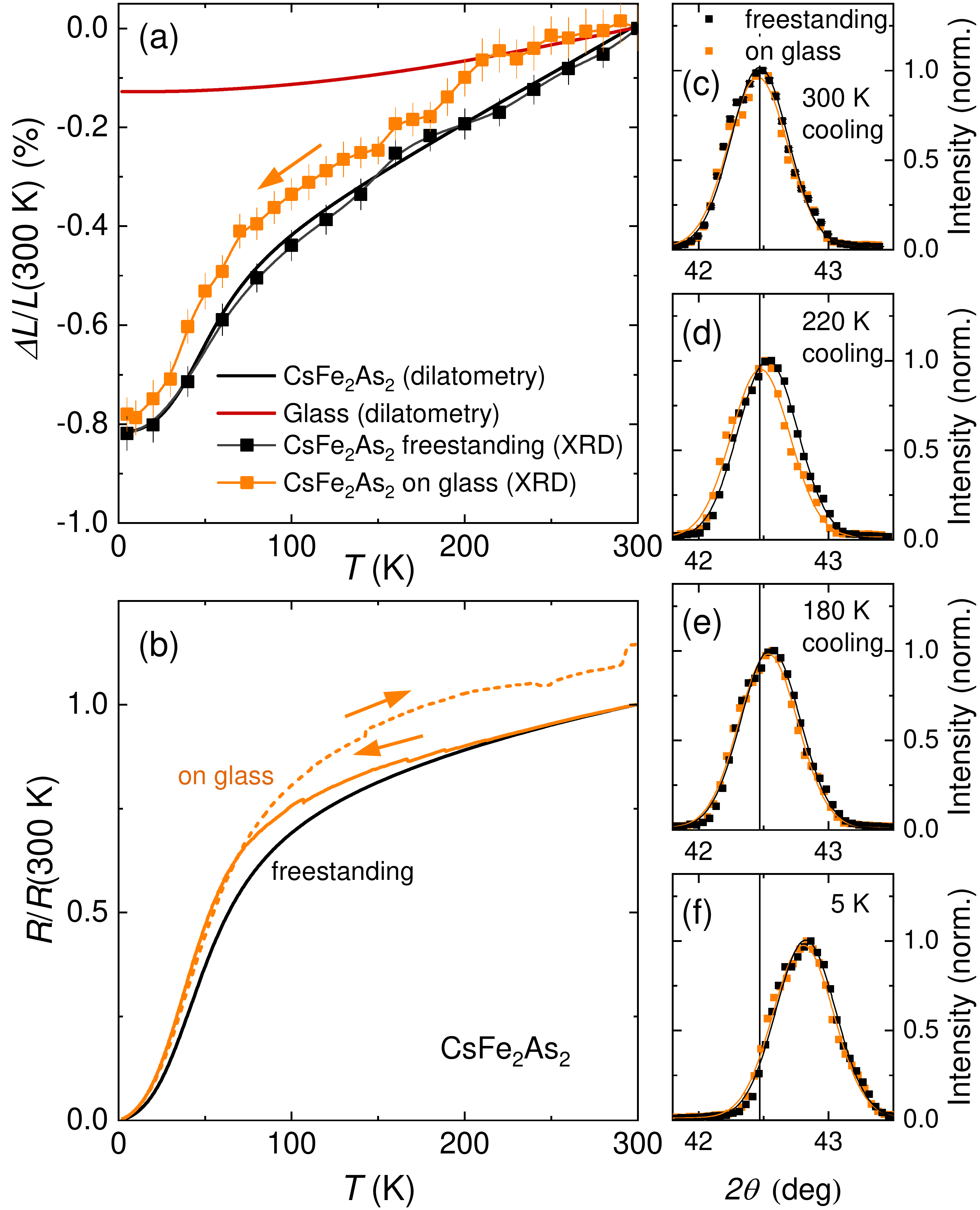}
\caption{(a) Solid lines: Temperature dependence of the change of sample length of a glass substrate\cite{Bohmer2017} (red) and freestanding CsFe$_2$As$_2$\cite{Hardy2016} (black), as measured in a capacitance dilatometer.
Squares: change of the $a$ lattice parameter of CsFe$_2$As$_2$ measured by x-ray diffraction, both freestanding (black squares) and glued to the glass substrate on a first cooling (red squares). 
(b) Resistance of a CsFe$_2$As$_2$ sample glued to glass on a first cooling (solid orange line) and on subsequent warming (dashed orange line).
This is compared with the resistance of a free standing crystal (black).
(c)-(f) X-ray intensity of longitudinal scans through the (4 0 0) Bragg reflection at indicated temperatures. Error bars are smaller
than the data points. Solid lines: single Gaussian fits from which the lattice parameters in (a) were extracted.
}
\label{fig:xray}
\end{figure}

Our results differ from previous reports of nematicity in CsFe$_2$As$_2$ and RbFe$_2$As$_2$, among them elastoresistance and STM results \cite{Ishida2019,Liu2019}. 
Most notably, our $m_{B_{1g}}$ and $m_{B_{2g}}$ elastoresistance coefficients are nearly temperature-independent, in contrast to the previous report\cite{Ishida2019}. Those data were obtained by gluing samples to the side of a piezoelectric stack, the technique most frequently used in the literature \cite{Chu2012}.
In order to understand the effects of gluing a CsFe$_2$As$_2$ sample to a substrate or sample holder of low thermal expansivity, such as a piezoelectric stack, we performed complementary resistance and XRD measurements on samples glued to thin glass plates. 
Fig. \ref{fig:xray} shows the temperature dependent length changes of the freestanding sample\cite{Hardy2016}, the glass substrate\cite{Bohmer2017} and the glued sample. 
Note that the sample length change measured directly using capacitance dilatometry and the length change inferred from the lattice parameters measured via XRD agree well for freestanding CsFe$_2$As$_2$, where both techniques can be used. 
Upon initial cooling, the length change of the glued sample follows closely the length change of the substrate down to $\sim$200 K [Fig. \ref{fig:xray}(a)]. Thus, the sample experiences elastic biaxial tension of up to $\sim$0.15$\%$. However, this tension is largely released below 200 K, demonstrating that sample and substrate are no longer perfectly bonded. 
In the same temperature range, the resistance shows discontinuous jumps [Fig. \ref{fig:xray}(b). 
Returning the sample to 300 K after cooling to base temperature reveals a large hysteresis. The behavior strongly suggests microscopic cracking of the sample, resulting in the strain release observed in XRD. 
This sample damage is a consequence of the unusually large thermal expansivity of CsFe$_2$As$_2$ and will affect any experiment in which the sample is glued to a substrate. 

In conclusion, we have shown that the elastoresistance of CsFe$_2$As$_2$ is dominated by the in-plane symmetric $A_{1g}$ strain component and that $A_{1g}$ strain sensitively induces modifications of the electronic entropy and correlations, which are reflected in the resistance. Furthermore, the absence of a divergent response in the symmetry-breaking $B_{1g}$ and $B_{2g}$ channels shows that CsFe$_2$As$_2$ is not near a nematic instability. Our results shed new light on the physics of the heavily hole-doped iron-based compounds close to an orbital-selective Mott transition. The experiment also highlights the importance of studying elastoresistance in all symmetry channels and demonstrates the method's versatility in the investigation of quantum materials beyond nematicity. 

Acknowledgements: We acknowledge support by the Helmholtz Association under contract number VH-NG-1242.
The contribution from M. M. was supported by the Karlsruhe Nano Micro Facility (KNMF).
We acknowledge valuable discussions with J. Schmalian, M. Kessler and K. Grube.

\end{document}